\documentclass{PoS}
\usepackage{amsmath}
\usepackage{dcolumn}

\title{%Contribution title%
Hadron properties at finite temperature and density with two-flavor Wilson fermions}

\ShortTitle{Hadron properties at finite temperature and density with two-flavor Wilson fermions}

\author{\speaker{Hideaki Iida}\\
        Mathematical Physics Lab., RIKEN Nishina Center, Wako, Saitama 351-0198, Japan\\
        E-mail: \email{hiida@riken.jp}}

\author{Yu Maezawa\\
        Mathematical Physics Lab., RIKEN Nishina Center, Wako, Saitama 351-0198, Japan\\
        }

\author{Koichi Yazaki\\
        Yukawa Institute for Theoretical Physics, Kyoto University, Kyoto 606-8502, Japan, and\\
        Mathematical Physics Lab., RIKEN Nishina Center, Wako, Saitama 351-0198, Japan
        }

\abstract{
Meson properties at finite temperature and density are studied in lattice QCD simulations 
 with two-flavor Wilson fermions. 
For this purpose, we investigate screening masses of mesons in pseudo-scalar (PS) and vector (V) channels. 
The simulations are performed on $16^3\times 4$ lattice along 
 the lines of constant physics  at $m_{\rm PS}/m_{\rm V}|_{T=0}=0.65$ and 0.80,
  where $m_{\rm PS}/m_{\rm V}|_{T=0}$ is a ratio of meson masses in 
   PS and V channels at $T=0$.  
A temperature range is $T/T_{\rm pc}=(0.8 - 4.0)$, where 
 $T_{\rm pc}$ is the pseudo-critical temperature. 
We find that the temperature dependence of the screening masses normalized by temperature, $M_0/T$, 
 shows notable structure around $T_{\rm pc}$, 
 and approach $2\pi$ at high temperature in both channels, 
  which is consistent with twice the thermal mass of a free quark 
   in high temperature limit. 
The screening masses at low density are also investigated 
 by using the Taylor expansion method with respect to the quark chemical potential. 
We find that the expansion coefficients in the leading order become positive in the temperature range,
 and thermal and density effect 
 on the meson screening-masses
  becomes apparent in the quark-gluon plasma phase.
The meson screening-masses are also compared with the gluon (Debye) screening masses
  at finite temperature and density.
}

\FullConference{The XXVIII International Symposium on Lattice Field Theory, Lattice2010\\
		June 14-19, 2010\\
		Villasimius, Italy}

\begin{document}

\section{Introduction}

Study of hadron properties at finite temperature and density is 
 important to understand behavior of quarks and gluons
  in hot and/or dense QCD medium.
In particular, mesons in the medium are expected to have abundant information
 about characteristic properties of QCD, 
  such as the deconfinement phase transition at finite temperature 
   and the partial chiral symmetry restoration at finite density \cite{Hatsudabon}.
In this article, we focus on screening masses of mesons in pseudo-scalar (PS) 
 and vector (V) channels calculated from the spatial correlation functions,
  and present current results of our group in lattice QCD simulations with two flavors of
   the Wilson-type quark action.
Using gauge configurations generated by WHOT-QCD Collaboration \cite{Maezawa2007,Ejiri2010},
 we calculate the meson correlation functions along the line of constant physics
  and extract temperature dependence of the meson screening-masses 
   from long spatial-distance behavior of the correlators.

We find that the screening masses in both channels show notable behavior 
  around pseudo-critical temperature $T_{\rm pc}$, where the transition occurs
   from the hadronic phase to the quark-gluon plasma phase.
On the other hand,
   these approach $2\pi T$ at high temperature, 
  which corresponds to twice the thermal mass of a non-interacting quark 
   in high temperature limit. 
We also investigate the meson screening-masses at low density 
 by using the Taylor expansion method with respect to the quark chemical potential.
We find that the leading order expansion 
coefficients become positive in the temperature range,
 and thermal and density effect 
 on the meson screening-masses
  becomes apparent in the quark-gluon plasma phase.
We also compare the meson screening-masses with the gluon (Debye) screening-masses
 obtained from Polyakov-line correlation functions, and find 
  characteristic difference in medium contributions between mesons and gluons.

The paper is organized as follows. 
In Sec.~2, we show the formalism to calculate the screening masses on the lattice. 
In Sec.~3, numerical results of the screening masses at finite temperature and zero density are discussed, 
and results of the screening masses at finite density in Sec.~4 using the Taylor expansion method. 
In Sec.~5, we compare the meson screening-masses
 with the gluon (Debye) screening masses. 
Section 6 is devoted to summarize the paper.

\section{Meson screening masses on lattice}

In order to extract the meson screening masses at finite temperature and density,
 we calculate an expectation value of 
the spatial correlation functions $G(x)$ of the mesons, 
\begin{eqnarray}
\langle G(x)\rangle \equiv \sum_{y,z,t} \langle M(x,y,z,t) M(0,0,0,0)^\dagger\rangle,
\end{eqnarray}
where $M(x,y,z,t)\equiv \bar q(x,y,z,t)\Gamma q(x,y,z,t)$ is the meson operator with  
$\Gamma$ denoting the gamma matrix, i.e., 
$\Gamma=\gamma_5$ for a pseudo-scalar (PS) meson 
 and $\Gamma=\gamma_\mu$ for a vector (V) meson. 
The correlator is summed over $y, z, t$, which means the zero-momentum projection 
for $y$ and $z$ directions and zero-energy projection for temporal direction. 

%%%%%%%%%%%%%%%%%%%%%%%
\def\simge{\mathrel{%
       \rlap{\raise 0.511ex \hbox{$>$}}{\lower 0.511ex \hbox{$\sim$}}}}
\def\simle{\mathrel{
       \rlap{\raise 0.511ex \hbox{$<$}}{\lower 0.511ex \hbox{$\sim$}}}}
%%%%%%%%%%%%%%%%%%%%%%%

In order to extract response to finite density,
 we apply the Taylor expansion method with respect to the quark chemical potential,
 $\mu \equiv(\mu_u+\mu_d)/2$, where 
 $\mu_u$ $(\mu_d)$ is the chemical potential for the $u$ ($d$) quark.
The Taylor expansion enables us to investigate meson properties at $\mu \neq 0$
 from the expectation values at $\mu = 0$.
Following Ref.~\cite{Choe2002c}, 
 we expand the expectation value of the operator $\cal O$ in powers of
 $\tilde\mu\equiv \mu/T$ as,
\begin{eqnarray}
\langle {\cal O} \rangle_\mu
%-----
&=&
\frac{\int {\cal D}U \ e^{-S_{\rm gluon}} ({\rm det} D(\mu))^2  {\cal O}{(\mu)}}
{\int {\cal D}U \ e^{-S_{\rm gluon}} ({\rm det} D(\mu))^2 } \nonumber \\
%-----
&=&
\frac{\int {\cal D}U \ e^{-S_{\rm gluon}} 
( \Delta + \dot \Delta \tilde\mu + \frac{1}{2} \ddot\Delta \tilde\mu^2
+ O(\tilde\mu^3))
({ \cal O} + \dot {\cal O} \tilde\mu + \frac{1}{2} \ddot{\cal O} \tilde\mu^2
+ O(\tilde\mu^3))
}
{\int {\cal D}U \ e^{-S_{\rm gluon}} 
( \Delta + \dot \Delta \tilde\mu + \frac{1}{2} \ddot\Delta \tilde\mu^2
+ O(\tilde\mu^3))
} \nonumber \\
%&  &\ \ \ \ \ \ \ \ \ \ \ \ \ \ \ \ \ \ \ \ \ \ \ \ \ \ \ \ \ \ \ \ \ \ \ \ ({\rm Hereafter \ we \ omit \ } `` \ |_{\mu=0} \ ".) \nonumber \\
%-----
&=&
\frac{
\langle
(
{\cal O} + \dot {\cal O}\tilde \mu +\frac{1}{2}\ddot {\cal O} \tilde\mu^2 + O(\tilde\mu^3)
)
(
1 + \frac{\dot \Delta}{\Delta}\tilde\mu + \frac{\ddot \Delta}{\Delta}\tilde\mu^2 + O(\tilde\mu^3)
)
\rangle
}
{
1 + \langle\frac{\dot \Delta}{\Delta}\rangle\tilde\mu + \frac{1}{2} \langle \frac{\ddot\Delta}{\Delta}
\rangle \tilde\mu^2 + O(\tilde\mu^3)
} \nonumber \\
%-----
 &\hspace{-1cm} =& \hspace{-0.5cm} \langle {\cal O}\rangle 
     + \left( \langle {\cal O} \frac{\dot\Delta}{\Delta} \rangle+ \langle {\cal \dot O}\rangle \right) \tilde \mu
     + \left(\langle \dot {\cal O} \frac{\dot \Delta}{\Delta} \rangle + \frac{1}{2}\langle\ddot {\cal O}\rangle 
     + \frac{1}{2}\langle{\cal O}\frac{\ddot\Delta}{\Delta}\rangle
     -  \frac{1}{2}\langle{\cal O}\rangle\langle \frac{\ddot\Delta}{\Delta}\rangle  \right) \tilde \mu^2 
     + O(\tilde \mu^3),
\label{eq1}
\end{eqnarray}
%#####
where $\Delta\equiv({\rm det} D (\mu))^2$ with $D(\mu)$ denoting   
the Dirac operator, and the dots on operators denote the derivatives with respect to $\tilde\mu$. 
Note that the expectation values in the right-hand-side of Eq.~(\ref{eq1})
  are calculated at $\mu=0$. 
We take ${\cal O}$ to be 
the meson correlator $G\equiv {\rm tr}(D^{-1}_{x0}(\mu_u)\Gamma D^{-1}_{0x}(\mu_d)\Gamma^\dagger)$ 
where ${\rm tr}$ denotes the trace with respect to
 color, spinor, and flavor indices.  
Then the Taylor expansion of the meson correlator is given by,
\begin{eqnarray}
\langle G \rangle_{\mu} &=& \langle G \rangle_0 + \langle G \rangle_1 \tilde\mu +
 \langle G \rangle_2 \tilde\mu^2 + O(\tilde\mu^3),
\end{eqnarray}
where the expansion coefficients become,
\begin{eqnarray}
\langle G \rangle_0 &=& 
 \langle {\rm tr}[D^{-1}_{x0}\Gamma \gamma_5 (D^{-1})^\dagger_{x0}\gamma_5\Gamma^\dagger]\rangle
,\\
\langle G \rangle_1 &=& 0
,\\
\langle G \rangle_2 &=& \langle G_{\rm opr} \rangle_2 + \langle G_{\rm det} \rangle_2
\label{eq:G2}
,\\
\langle G_{\rm opr}\rangle_2 &\equiv& \langle \dot G \frac{\dot \Delta}{\Delta} \rangle + \frac{1}{2}\langle\ddot G \rangle
 \nonumber\\
&=&
2\langle{\rm Re}{\rm tr}[(D^{-1}\dot D D^{-1}\dot D D^{-1})_{x0}\Gamma \gamma_5 (D^{-1})^\dagger_{x0}\gamma_5\Gamma^\dagger]\rangle\nonumber\\
  &&-\langle{\rm Re}{\rm tr}[(D^{-1}\ddot D D^{-1})_{x0}\Gamma \gamma_5 (D^{-1})^\dagger_{x0}\gamma_5 \Gamma^\dagger]\rangle \nonumber \\
  &&-\langle{\rm Re}{\rm tr}[(D^{-1}\dot D D^{-1})_{x0}\Gamma\gamma_5 (D^{-1}\dot D D^{-1})^\dagger_{x0}\gamma_5 \Gamma^\dagger]\rangle \nonumber \\
  &&+4\langle {\rm Im}{\rm tr} [(D^{-1}\dot D D^{-1})_{x0}\Gamma\gamma_5 (D^{-1})^\dagger_{x0}\gamma_5\Gamma^\dagger] \cdot {\rm Im} {\rm Tr}(D^{-1}\dot D)\rangle
, \nonumber\\
 \langle G_{\rm det}\rangle_2 &\equiv& \frac{1}{2}\langle G \frac{\ddot\Delta}{\Delta}\rangle
     -  \frac{1}{2}\langle G \rangle\langle \frac{\ddot\Delta}{\Delta}\rangle
\nonumber\\
 &=& {\rm Re}\{ \langle {\rm tr}[D^{-1}_{x0}\Gamma\gamma_5 (D^{-1})^\dagger_{x0}\gamma_5\Gamma^\dagger]
    (2({\rm Tr}(D^{-1}\dot D))^2-{\rm Tr}(D^{-1}\dot D D^{-1} \dot D) +{\rm Tr}(D^{-1}\ddot D))\rangle \nonumber \\
   && -\langle {\rm tr}[D^{-1}_{x0}\Gamma\gamma_5 (D^{-1})^\dagger_{x0}\gamma_5\Gamma^\dagger]\rangle
    \langle 2({\rm Tr}(D^{-1}\dot D))^2-{\rm Tr}(D^{-1}\dot D D^{-1} \dot D) +{\rm Tr}(D^{-1}\ddot D)\rangle \}
   , \nonumber
\end{eqnarray}
where ${\rm Tr}$ denotes the 
trace including space-time indices in addition to those for ${\rm tr}$. 
We have divided the second derivatives into two parts: 
 $\langle G_{\rm opr} \rangle_2$ and $\langle G_{\rm det} \rangle_2$,
  where the former includes the derivatives of the operator $G$,
%   (first two terms in the coefficient of $\tilde\mu^2$ of Eq.~(\ref{eq1})), 
  and the latter consists only of the derivatives of the quark determinant.
%  (last two terms in the coefficient of $\tilde\mu^2$ of Eq.~(\ref{eq1})). 
Note that the meson correlator does not have the odd orders in the Taylor expansion
 since it is symmetric under $\mu \rightarrow - \mu$, namely the meson correlator is 
  invariant under the charge conjugation.
For the calculation of the trace with respect to spatial indices,
 we apply the random noise method with 100 sets of U(1) random numbers. 

In order to study the screening effect, 
 we fit the meson correlator by the following form,
\begin{eqnarray}
\langle G (x) \rangle_{(\mu, T)}
 = A(\mu,T) \left[ e^{-M(\mu,T) x}+e^{-M(\mu,T)(L-x)} \right]
,
\end{eqnarray}
where $L$ is the spatial lattice size, 
and we assume that contributions of finite $\mu$ appear 
 only in the coupling factor ($A$) and the meson screening mass ($M$). 
We also assume that $A(\mu,T)$ and $M(\mu,T)$ are also expressed 
 as power series in $\tilde\mu$,
\begin{eqnarray}
A(\mu,T) &=& A_0 + A_2 \tilde\mu^2 + O(\tilde\mu^4) , \\
M(\mu,T) &=& M_0 + M_2 \tilde\mu^2 + O(\tilde\mu^4) .
\end{eqnarray}
By comparing both sides of Eq.~(2.8) at each order of $\tilde\mu$, we obtain,
\begin{eqnarray}
\langle G(x)\rangle_0 &=& A_0 \left( e^{-M_0 x}+e^{-M_0 (L- x)} \right)
\label{fit_0th}
,\\
\frac{\langle G(x) \rangle_2}{\langle G(x)\rangle_0} 
 &=& \frac{A_2}{A_0}+ M_2
\left\{ \left(x-\frac{L}{2}\right)\tanh\left[M_0 \left( x-\frac{L}{2}\right)\right]
-\frac{L}{2}\right\}.
\label{fit_second}
\end{eqnarray}
We extract the screening masses
by fitting the meson correlators for each temperature with the expressions
Eqs.~(\ref{fit_0th}) and (\ref{fit_second}) at large distance.

\section{Numerical simulations}

Simulation setup is the following. 
We utilize the gauge configurations generated by WHOT-QCD Collaboration  
 on $16^3\times 4$ lattice with the renormalization-group improved Iwasaki gauge action and 
   $N_f=2$ clover-improved Wilson quark action \cite{Maezawa2007,Ejiri2010}.  
The simulations have been performed  along the line of constant physics 
 corresponding to the PS and V meson mass ratio, 
  $m_{\rm PS}/ m_{\rm V}|_{T=0}=0.65$ and 0.80 at $T=0$.
The temperature range for $m_{\rm PS}/ m_{\rm V}|_{T=0}=0.65$ (0.80) is 
$T/T_{\rm pc}=$0.82--4.0 (0.76--3.0), where $T_{\rm pc}$ is the pseudo-critical temperature
 for the transition from hadronic phase to quark-gluon plasma phase. 
The number of configurations we use is 100 for each temperature and quark mass. 
% The errorbars are estimated by jackknife method.  

\subsection{Screening masses at finite temperature and zero density}

Figure 1(a) shows the meson screening-masses normalized by temperature, $M_0/T$,
 in PS channel as a function of $T/T_{\rm pc}$. 
The circle (triangle) points correspond 
 to the results at $m_{\rm PS}/m_{\rm V}|_{T=0}=0.65$ (0.80). 
The same figure in V channel is shown in Fig.~1(b).
We can see a concave structure around $T_{\rm pc}$,
 that is, when temperature increases, 
  $M_0 /T$ decreases below $T_{\rm pc}$, whereas it increases above $T_{\rm pc}$. 
This implies that the screening masses $M_0$ stay constant below $T_{\rm pc}$, 
whereas they monotonically increase above $T_{\rm pc}$.
Namely, thermal effect on the meson screening-masses
 becomes apparent in the quark-gluon plasma phase.
At high temperature, $M_0/T$ 
  converges to a constant value of $2 \pi$,
 which implies that 
 the meson becomes a weakly interacting pair of a quark and an anti quark, 
 each carrying the thermal mass $\pi T$.
%Figure 1(b) shows the meson screening-masses $M_0/T$ in V channel.
%We can see that the behavior is similar to that in PS channel, 
% although the concave structure is less conspicuous in V channel. 
We also find clear quark-mass dependence, i.e., magnitude of the screening masses 
  with lighter quark mass ($m_{\rm PS}/m_{\rm V}|_{T=0}=0.65$)
 becomes smaller than that with heavier quark mass ($m_{\rm PS}/m_{\rm V}|_{T=0}=0.80$)
    in both channels, similarly to the ordinary meson mass measured by temporal correlation. 
This implies that the screening of the meson spatial correlation 
 becomes weak when the quark mass becomes small.

\begin{figure}[t]
\begin{minipage}{8.5cm}
\hspace{-0.5cm}
\includegraphics[width=0.97\textwidth]{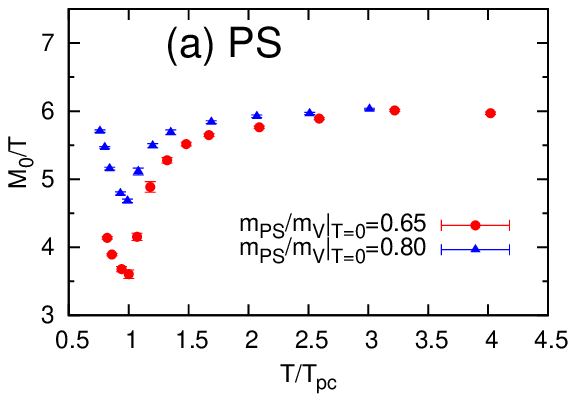}
\label{fig1}
\end{minipage}
\hspace{-1.5cm}
\begin{minipage}{8.5cm}
\includegraphics[width=0.97\textwidth]{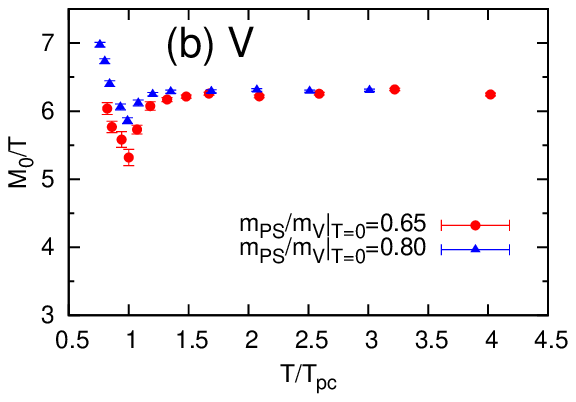}
\label{fig1}
\end{minipage}
\caption{Meson screening masses $M_0/T$ in the PS channel (a) and V channel (b)
 as a function of temperature.}
\end{figure}

\if0
%#####
We also show a ratio of the screening masses, $M_0^{({\rm PS})}/M_0^{({\rm V})}$,
 in Fig.~2(a) and 2(b) at $m_{\rm PS}/m_{\rm V}|_{T=0}=0.65$ and 0.80, respectively.
It is expected that the ratio approaches to the value of $m_{\rm PS}/m_{\rm V}|_{T=0}$
 at low temperature, and it goes to unity at high temperature where
  the mesons consist of the weakly interacting quark-antiquark pairs.
We can see the expected behavior, that is, 
 $M_0^{({\rm PS})}/M_0^{({\rm V})} \sim 0.65$ (0.80)
 in the left (right) panel of Fig.~2  below $T_{\rm pc}$,
  whereas these go to unity above $T_{\rm pc}$.
\fi

The temperature dependence of the meson screening masses is
 consistent with that calculated in the staggered-type quark action \cite{Cheng2010}.

\if0
%###############

\begin{figure}[t]
\begin{minipage}{8.7cm}
\hspace{-0.5cm}
\includegraphics[width=0.97\textwidth]{./PSovV_mPSovmV0.65_2.eps}
\label{fig1}
\end{minipage}
\hspace{-1.7cm}
\begin{minipage}{8.7cm}
\includegraphics[width=0.97\textwidth]{./PSovV_mPSovmV0.80_2.eps}
\label{fig1}
\end{minipage}
\caption{Ratio of the screening masses, $M_0^{({\rm PS})}/M_0^{({\rm V})}$, 
 at $m_{\rm PS}/m_{\rm V}|_{T=0}=0.65$ (a) and 0.80 (b) as a function of temperature.}
\end{figure}
%###############
\fi

\subsection{Screening masses at finite density}

In this section, we investigate properties of the screening masses at finite density 
 by calculating the second response of the screening masses $M_2/T$ 
  to the quark chemical potential $\mu$ via the Taylor expansion method.
Figure 2(a) and 2(b) show temperature dependence 
 of $M_2/T$  in PS and V channels, respectively, 
 with $m_{\rm PS}/m_{\rm V}|_{T=0}=0.65$ (circle points) and 0.80 (triangle points).  
We find that $M_2/T$ is always positive 
 in the temperature range we have explored, 
  which implies that the screening masses increase 
   in the leading order contribution of $\mu$. 
$M_2/T$ increases rapidly at $T_{\rm pc}$ which 
 means that density effect on the meson screening-masses
 becomes significant in the quark-gluon plasma phase.
We also found from the simulations that
 main contribution to the rapid increases of $M_2/T$ comes from
  the $\langle G_{\rm opr}\rangle_2$ in Eq.~(\ref{eq:G2})
   which consists of the derivatives of the operator.
We can also see the quark-mass dependence in the PS channel; 
 the magnitude of $M_2/T$ 
  with lighter quark mass ($m_{\rm PS}/m_{\rm V}|_{T=0}=0.65$)
   is slightly smaller than that with heavier quark mass ($m_{\rm PS}/m_{\rm V}|_{T=0}=0.80$).
This is similar to the quark-mass dependence of $M_0/T$,
 but different from the results of $M_2$ calculated 
  in the staggered-type quark action \cite{Choe2002c}.
This difference should be further
  investigated in the simulations with
   smaller quark mass and larger spacial volume.

%We also examine the contributions of 
%the operator part $\langle\ddot G_{\rm opr}\rangle$ and the determinant part $\langle \ddot G_{\rm det}\rangle$ 
%in Eq.~(\ref{corr_second}) to the second derivatives, 
%and find that the rapid increase of $Td^2M/d\mu^2$ is mainly due to the operator part 
%and the contribution of the determinant part is small, although the results are not shown in the figures. 
%As for the quark mass dependence, the derivatives are about the same for two quark masses in V channel, 
%while they are larger for the larger quark mass in PS channel. 
%{\bf 
%The quark mass dependence of second derivatives 
%seems to be opposite for the results in PS channel obtained by staggered fermions \cite{Choe2002c}. 
%Note that the fit in second order is not so good at high temperature in our calculation, 
%which may be originated from the finite volume effect. 
%Also note that the fit range dependence of $Td^2M/d\mu^2$ is rather large. 
%}

 \begin{figure}[t]
\begin{minipage}{8.7cm}
\hspace{-0.5cm}
\includegraphics[width=.95\textwidth]{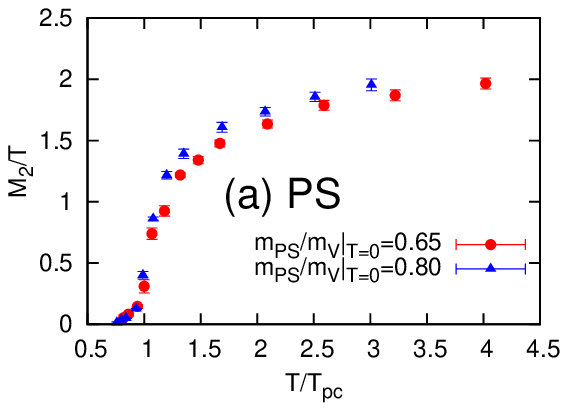}
\label{fig1}
\end{minipage}
\hspace{-1.7cm}
\begin{minipage}{8.7cm}
\includegraphics[width=.95\textwidth]{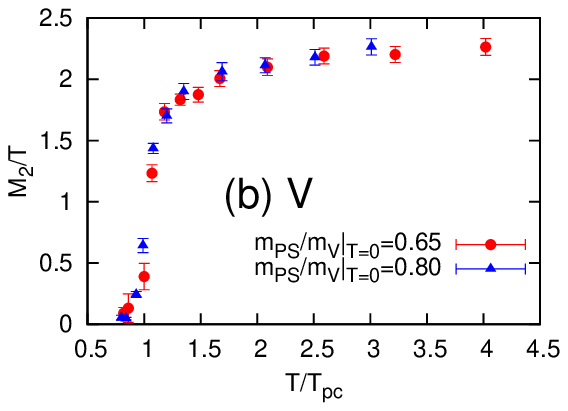}
\label{fig1}
\end{minipage}
\caption{Second response of the screening masses $M_2/T$ with respect to $\tilde\mu$
 as a function of $T/T_{\rm pc}$ for PS channel (a) and V channel (b).}
\end{figure}

\section{Comparison with gluon screening-masses}

Let us compare the screening masses of the PS mesons with these of the gluons.
The response to the quark chemical potential $\mu$ should be different
 for fermionic objects and gluonic ones, because
  the former (such as the meson screening-mass) 
   have direct coupling with $\mu$ described by $\langle G_{\rm opr} \rangle_2$
    while the latter (such as the gluon screening-mass) have only indirect coupling
     via the dynamical quark loops in the medium.
Therefore, it is important to compare these two screening masses.
The gluon screening-mass (so called Debye screening mass) has been
  studied from the Polyakov-line correlator based on the Taylor expansion method
   using the same gauge configurations \cite{Maezawa2007,Ejiri2010}. 
The Debye screening mass is also expressed as power series of $\tilde\mu$,
\begin{eqnarray}
m_D(\mu) = m_{D,0} + m_{D,2} \tilde\mu^2 + O(\tilde\mu^4).
\end{eqnarray}

Figure 3 shows the meson (a) and Debye (b) screening-masses
 at $m_{PS}/m_V|_{T=0}=0.65$, where the circle (triangle) plots show
 $M_0/T$ and $m_{D,0}/T$ ($M_2/T$ and $m_{D,2}/T$), respectively.
We find characteristic difference in the temperature dependence,
 i.e. the meson (Debye) screening-masses increase (decrease) when
  temperature increases above $T_{\rm pc}$.
This is related to the fact that, at high temperature limit,
 $M_0$ goes to $2 \pi T$ which is twice the thermal mass of a free quark, 
  whereas $m_{D,0}$ is proportional to the running coupling and
   goes to zero according to the prediction of the thermal perturbation theory.
We also find that the ratio of the meson screening-masses, $M_2/M_0$, is
 larger than that of the Debye screening-masses, $m_{D,2}/m_{D,0}$.
This means that the response of the fermionic object to $\mu$
 is larger than that of the gluonic object
  due to the difference in their ways of coupling with $\mu$ mentioned above.

\begin{figure}[t]
\begin{minipage}{8.7cm}
\includegraphics[width=.9\textwidth]{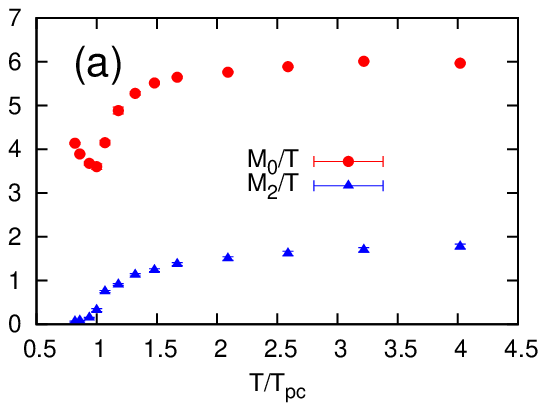}
\label{fig1}
\end{minipage}
\hspace{-1.2cm}
\begin{minipage}{8.7cm}
\includegraphics[width=.9\textwidth]{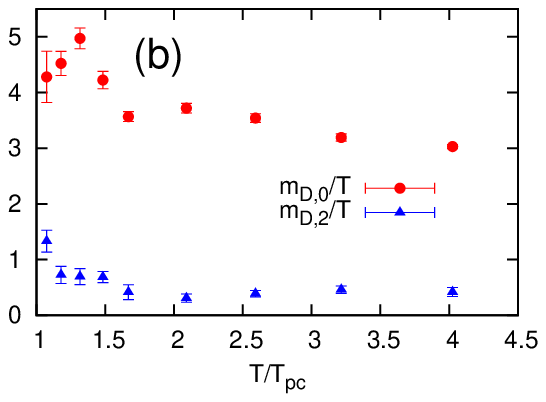}
\label{fig1}
\end{minipage}
\caption{(a) meson screening-masses in PS channel, (b) gluon (Debye) screening-masses
 at $m_{PS}/m_V|_{T=0}=0.65$ as a function of $T/T_{\rm pc}$.}
\end{figure}

\section{Summary}

We have studied the meson screening-masses in PS and V channel
  at finite temperature and density in lattice QCD simulations
   with two-flavor Wilson fermions.
The simulations have been performed along the line of constant physics at
 $m_{PS}/m_V|_{T=0}=0.65$ and 0.80 with the temperature range of 
 $T/T_{\rm pc} = 0.82$--4.0 and 0.76--3.0, respectively. 
On the basis of the Taylor expansion method, we have calculated temperature dependence
  of the leading order term of the screening masses ($M_0$) and 
   the second response to the quark chemical potential $\mu$ ($M_2$).
We have found that $M_0/T$ shows a concave structure around $T_{\rm pc}$
 and goes to $2 \pi$ at high temperature, which corresponds to 
  twice the thermal mass of a free quark. 
From the quark-mass dependence, we have seen that
  the screening effect on the meson correlator   
 becomes weak when the quark mass becomes small.

\if0
The ratio of the screening mass $M_0^{({\rm PS})}/M_0^{({\rm V})}$
 shows the expected behavior, i.e. it approaches to the value of
  $m_{\rm PS}/m_{\rm V}|_{T=0}$ at low temperature,
 and it goes to unity at high temperature where
  the mesons consist of the weakly interacting quark-antiquark pairs.
\fi

The second response $M_2/T$ is always positive 
 in the temperature range we have explored, 
  which implies that the screening masses increase 
   in the leading order contribution of $\mu$. 
We have also found that thermal and density effect 
 on the meson screening-masses
  becomes significant in the quark-gluon plasma phase.

We have also compared the meson screening-masses 
 with the gluon (Debye) screening-masses, and found 
  characteristic difference of the temperature dependence,
 i.e. the meson (Debye) screening-masses increase (decrease) when
  temperature increases.
We have also seen that the response to $\mu$ is different
 between the meson and Debye screening-masses, which reflects the fact that 
   the fermionic object has direct coupling with $\mu$,
   whereas the gluonic object has only indirect coupling
     via the dynamical quark loops in the medium.

\acknowledgments
We thank WHOT-QCD Collaboration for providing us with the gauge configurations. 
The calculations were performed by using the RIKEN Integrated Cluster of Clusters (RICC) facility.

%\bibliographystyle{./JHEP-2.bst}
%\bibliography{/users/iida/Documents/JabRef_files/bib_files/lattice2010_proceedings}  
\providecommand{\href}[2]{#2}\begingroup\raggedright\endgroup

\end{document}